\documentclass{iau} 
\usepackage{graphicx}

\title[Zirconium abundance and red giants] 
{Abundance of zirconium in the atmospheres of red giants in Galactic globular cluster 47~Tuc}

\author[Edgaras Kolomiecas, Vidas Dobrovolskas and Ar\={u}nas Ku\v{c}inskas]   
{Edgaras Kolomiecas$^1$, Vidas Dobrovolskas$^1$ and Ar\={u}nas Ku\v{c}inskas$^1$
}

\affiliation{
$^1$Institute of Theoretical Physics and Astronomy, Vilnius University\\ Saul\.{e}tekio av. 3, LT-10222, Vilnius, Lithuania \\ email: {\tt edgaras.kolomiecas@ff.vu.lt, vidas.dobrovolskas@tfai.vu.lt, arunas.kucinskas@tfai.vu.lt} \\[\affilskip]
}

\pubyear{2019}
\volume{351}  
\jname{Star Clusters: From the Milky Way to the Early Universe}
\editors{A. Bragaglia, M.B. Davies, A. Sills \& E. Vesperini, eds.}
\begin{document}

\maketitle

\begin{abstract}
We determined zirconium abundance in the atmospheres of 327 red giant branch (RGB) stars in the globular cluster 47~Tuc. The 1D~LTE abundances were obtained from the archival VLT~GIRAFFE spectra, using 1D hydrostatic {\tt ATLAS9} stellar model atmospheres and synthetic Zr~I line profiles computed with the {\tt SYNTHE} package. The average zirconium abundance determined in the sample of RGB stars, $\langle {\rm [Zr/Fe]} \rangle = +0.38 \pm 0.12$, agrees well with zirconium abundances obtained at this metallicity in the Galactic field stars, as well as with those observed in other Galactic globular clusters.
\end{abstract}

\firstsection 

\section{Introduction}

Galactic globular clusters (GGCs) were long believed to be simple stellar populations (SSPs). This paradigm has changed dramatically during the last few decades when photometric and spectroscopic studies revealed that almost all GGCs are built of multiple (at least two) stellar populations (e.g., \cite[Bastian \& Lardo 2018]{BastianLardo2018}). It is currently thought that stars that belong to the first population retained chemical composition of the cloud from which the cluster itself had formed, whereas stars in the second population have been enriched in certain light elements (e.g., N, Na) and depleted in others (e.g., C, O; see, e.g., \cite{Carretta2009}). A number of evolutionary scenarios of the GGCs have been proposed in order to explain the observed trends, such as pollution by massive asymptotic giant branch (AGB) stars (\cite[Ventura et al. 2001]{Ventura2001}), fast rotating massive stars \mbox{(Decressin et al. 2007a,b)}, or early disk accretion scenario (\cite[Bastian et al. 2013]{Bastian2013}). Unfortunately, none of the proposed scenarios are able to explain all observed chemical and kinematical properties of globular cluster populations simultaneously.

One may anticipate that additional information about the possible polluters could be obtained by analysing abundances of $s$-process elements. Indeed, in their recent study \cite[Gratton et al. (2013)]{Gratton2013} have suggested the existence of mild Ba--Na correlation in 47~Tuc. However, it is unclear whether such correlations may also exist in case of other $s$-process elements. Potentially, this may allow to constrain the mass range of the possible polluters, because the ratio of light $s$-elements (e.g., Sr, Zr) to heavy $s$-process elements (e.g., La, Ba) synthesized in AGB stars depends strongly on stellar mass (Straniero et al. 2014). Therefore, in this study we determined zirconium abundance in the Galactic globular cluster 47~Tuc with the aim of better understanding nucleosynthesis of $s$-process elements in this GGC.

\section{Observations and data reduction}


To determine zirconium abundance, we utilized high-resolution spectra of 327 red giant branch (RGB) stars in 47~Tuc. This is the largest stellar sample in which zirconium abundance has been determined in any GGC to date (Fig.~1). Abundance analysis was based on the archival spectra of RGB stars that were obtained with the GIRAFFE spectrograph mounted on the ESO VLT UT2 telescope. The spectra were obtained during three observing programmes: 072.D-0777, 073.D-0211 and 088.D-0026 (HR13, \textit{R} = 26\,400, typical $S/N \approx 75$).

Effective temperatures and surface gravities of the sample stars were determined from photometry, microturbulence velocities were obtained from Fe~I lines. For the effective temperature determination, we used photometry from \cite[Bergbusch \& Stetson (2009)]{BS2009} and colour-effective temperature calibrations of \cite[Ram\'{\i}rez \& Mel\'{e}ndez (2005)]{RM2005}. Surface gravities were obtained using a relation between surface gravity, stellar mass, luminosity, and effective temperature. We assumed a fixed mass of 0.9\,M$_{\odot}$ for all RGB stars, as determined using Yonsei-Yale isochrone of 12~Gyr and cluster metallicity of [Fe/H] = $-$0.68. We used a relation between the bolometric correction, effective temperature, and metallicity from \cite[Alonso et al. (1999)]{Alonso1999} to estimate stellar luminosity.

Two Zr~I lines were used in the abundance determination, with their central wavelengths located at 613.4585\,nm and 614.3252\,nm (atomic parameters of Zr~I lines used in our work are summarized in Table~1). Line equivalent widths were measured using the {\tt DECH20T} package, by fitting Gaussian profiles to the observed spectral line profiles. Stellar model atmospheres were computed using the {\tt ATLAS9} code and were further employed to derive 1D~LTE zirconium abundances with the {\tt WIDTH9} package.

\begin{table}[h]
	\centering
	\caption{Atomic parameters of Zr~I  lines used in the abundance determination. Natural, Stark, and van der Waals broadening constants are given in the last three columns, respectively. }
	\begin{tabular}{c c c c c c c}
		\hline
		Element   & $\lambda$, nm  &$\chi$, eV & log\textit{gf} & log$\gamma$$_{rad}$ & log$\frac{\gamma_{4}}{N_{e}}$ &  log$\frac{\gamma_{6}}{N_{H}}$\\
		\hline
		Zr~I	&	613.4585	&	0.000	&	$-$1.277	&	7.77	&	$-$5.70	&	$-$7.79 \\
		Zr~I	&	614.3252	&	0.071	&	$-$1.097	&	7.77	&	$-$5.69	&	$-$7.79 \\

		\hline
	\end{tabular}
\end{table}

\section{Results}

\begin{figure}[!h]
	\begin{center}
		\includegraphics[width=4.0in]{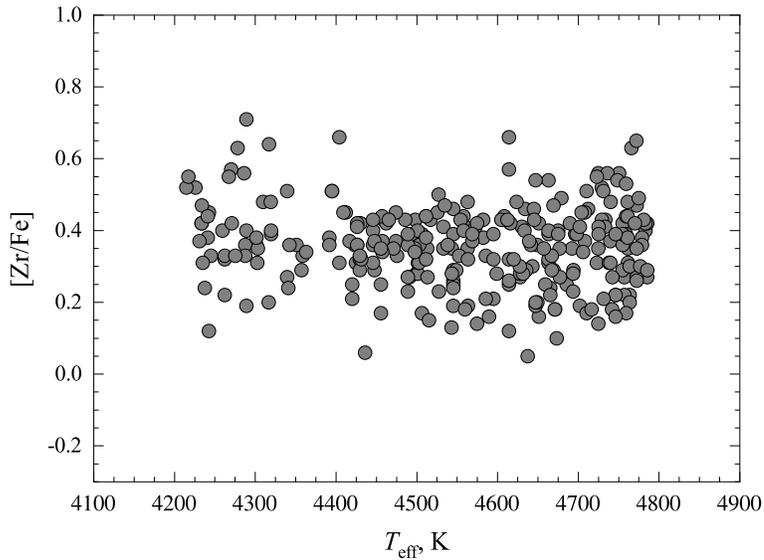} 
		\caption{Zirconium abundance in the RGB stars of 47~Tuc, plotted as a function of effective temperature.}
		\label{fig1}
	\end{center}
\end{figure}

\begin{figure}[!h]
	\begin{center}
		\includegraphics[width=4.0in]{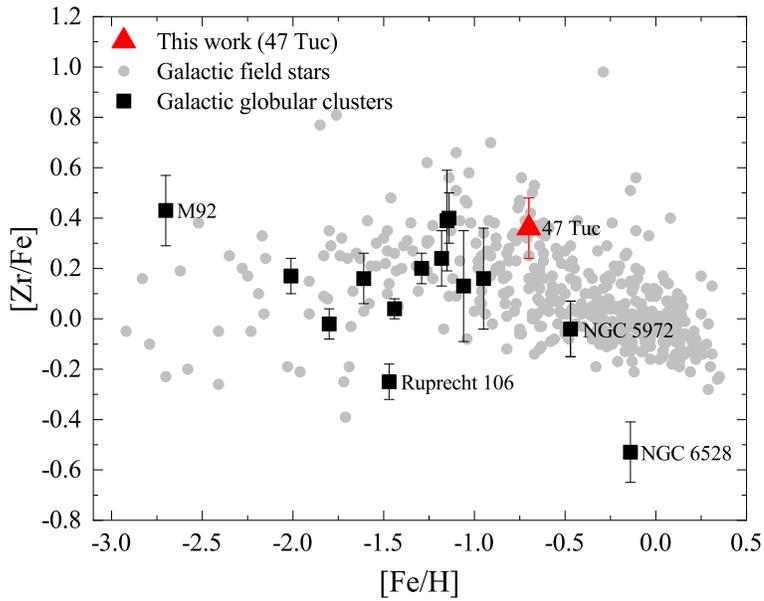} 
		\caption{Abundance of zirconium in the individual GGCs (black squares) and Galactic field stars (grey circles), plotted in the [Zr/Fe]--[Fe/H] plane. Red triangle is [Zr/Fe] ratio in 47~Tuc determined in this study. Error bars show the spread of [Zr/Fe] ratios in the individual stars of a given cluster (standard deviation due to star-to-star abundance scatter).}
		\label{fig2}
	\end{center}
\end{figure}

The mean zirconium abundance that we obtained in a sample of 327 RGB stars in 47~Tuc is [Zr/Fe]= $+$0.38 $\pm$ 0.12 (the error here is standard deviation due to star-to-star abundance variation). This is so far the largest sample of RGB stars analysed in this cluster for zirconium abundance. The obtained result (shown as red triangle in Fig. 2, together with zirconium abundances in the Galactic field stars and other GGCs taken from literature) suggests that nucleosynthesis of zirconium in the GGCs and Galactic field stars has proceeded along similar pathways.

\end{document}